\documentclass[sigconf, authorversion,nonacm]{acmart}
\AtBeginDocument{%
  \providecommand\BibTeX{{%
    \normalfont B\kern-0.5em{\scshape i\kern-0.25em b}\kern-0.8em\TeX}}}

\setcopyright{acmlicensed}
\copyrightyear{2024}
\acmYear{2024}
\acmDOI{XXXXXXX.XXXXXXX}
\usepackage{subcaption}

\usepackage{geometry}

\usepackage{url}
\usepackage{cleveref}

\begin{document}

\title{Characterizing and modeling harms from interactions with design patterns in AI interfaces}

\author{Lujain Ibrahim}

\email{lujain.ibrahim@oii.ox.ac.uk}
\affiliation{
  \institution{Oxford Internet Institute}
  \city{Oxford}
  \country{UK}
}
\authornote{Corresponding author: lujain.ibrahim@oii.ox.ac.uk}

\author{Luc Rocher}
\email{luc.rocher@oii.ox.ac.uk}
\affiliation{%
  \institution{Oxford Internet Institute}
  \city{Oxford}
  \country{UK}
}
\authornote{These authors contributed equally to this work and share senior authorship.}

\author{Ana Valdivia}
\email{ana.valdivia@oii.ox.ac.uk}
\affiliation{%
  \institution{Oxford Internet Institute}
  \city{Oxford}
  \country{UK}
}
\authornotemark[2]
\renewcommand{\shortauthors}{Ibrahim et al.}

\begin{abstract}
 The proliferation of applications using artificial intelligence (AI) systems has led to a growing number of users interacting with these systems through sophisticated interfaces. Human-computer interaction research has long shown that interfaces shape both user behavior and user perception of technical capabilities and risks. Yet, practitioners and researchers evaluating the social and ethical risks of AI systems tend to overlook the impact of anthropomorphic, deceptive, and immersive interfaces on human-AI interactions. Here, we argue that design features of interfaces with adaptive AI systems can have cascading impacts, driven by feedback loops, which extend beyond those previously considered. We first conduct a scoping review of AI interface designs and their negative impact to extract salient themes of potentially harmful design patterns in AI interfaces. Then, we propose \textbf{D}esign-\textbf{E}nhanced \textbf{C}ontrol of \textbf{AI} systems (DECAI), a conceptual model to structure and facilitate impact assessments of AI interface designs. DECAI draws on principles from control systems theory---a theory for the analysis and design of dynamic physical systems---to dissect the role of the interface in human-AI systems. Through two case studies on recommendation systems and conversational language model systems, we show how DECAI can be used to evaluate AI interface designs. 
\end{abstract}


\maketitle

\section{Introduction}
Decades-long trends have shown that the success and popularity of a technology depend on the usability, accessibility, and user-friendliness of its interface \cite{godwin2008mobile, knutson1997effect}. This trend is also evident in the development of new applications with artificial intelligence (AI) systems; scholars have suggested that TikTok's broad popularity should be attributed not only to its recommendation algorithm but also to its vertical video design, which, among other things, makes bad recommendations less noticeable \cite{Narayanan_2022, striano2023alert}. Analyses of ChatGPT's virality regularly point to its minimalistic and unobtrusive user interface as being a key driver of its success \cite{THE_CONVERSATION_2023, Ramlochan_2023}. 

Interfaces are key drivers of the adoption of AI applications, yet their impact is understudied in the evaluation of harms and risks from AI systems. For instance, audits of algorithmic decision-making aids investigate system fairness by observing how changes in model inputs influence outputs \cite{bandy2021problematic}; studies of polarization driven by recommendation algorithms use sock puppet accounts to simulate different engagement patterns and analyze the resulting recommendations \cite{haroon2023auditing}; safety benchmarks in natural language processing and generation evaluate how models themselves could avoid generating undesirable content \cite{chang2023survey}. While all of these evaluations are critical indicators of potential downstream harms, they do not account for the design and structure of the user interface through which the majority of people receive AI systems post-deployment.

Research in human-computer interaction (HCI) and science and technology studies has shown that technology cannot be studied by abstracting out interfaces \cite{ruijten2018enhancing, coskun2005impacts}. A notable area of research in this space has examined unethical interface designs, known as \textit{dark patterns}; studies on dark patterns continue to reveal the prevalence of digital designs that steer users to complete, often detrimental, actions that they would not necessarily complete otherwise \cite{gray2018dark, mathur2021makes}. Interfaces do not only facilitate such autonomy-undermining influence on user behavior, but they also shape user perceptions of technologies, their capabilities, and their risks \cite{satzinger1998user}. For example, consider how explanations of AI predictions have been integrated into interfaces that mediate human interactions with AI decision-making aids. While intended to improve fairness, research has shown that how these explanations are presented in interfaces can lead decision-makers to place unwarranted trust in biased AI predictions \cite{ibrahim2023explanations, green2019disparate}.
 
Critically, in AI applications, user interfaces are also sites of data collection for increasingly \textit{adaptive} AI systems. Every time a user interacts with an adaptive system, they supply it with new information that influences that system's future outcomes \cite{matias2023humans, mansoury2020feedback}. This creates \textit{feedback loops} of human-AI behavior, that shape outcomes without additional involvement from system developers \cite{matias2022impact}. For instance, in social media platforms, user engagement with certain types of content can lead the recommendation algorithm to prioritize similar content, perpetuating a cycle of exposure and interaction \cite{loosen2016caught}. The design of AI interfaces, therefore, directly influences these temporal dynamics, as interfaces mediate how users receive and respond to model output over time. As such, interface designs may contribute to cascading harms, both at the individual level, such as addictive and extractive usage, and at the societal level, like the spread of misinformation.

To better understand the ethical and social risks of AI systems post-deployment, research must be able to scrutinize AI interface designs. Here, we bridge HCI research on interfaces with the scholarship on AI's harms and risks by developing a conceptual model to aid in the assessment of AI interface design choices. We call this model \textbf{D}esign-\textbf{E}nhanced \textbf{C}ontrol of \textbf{AI} systems (DECAI). DECAI's design synthesizes the results of a scoping review of harmful AI interface designs (Section~\ref{sec:review}) with principles from control systems theory---a theory for the design and analysis of dynamic physical systems. DECAI breaks down the role of the interface in processing system input and presenting system output, providing a structure for generating testable hypotheses for evaluating AI interface designs (Section~\ref{sec:decai}). We show how DECAI can be used as a starting point to analyze the impact of design features on different user groups using two case studies on high-stakes adaptive AI systems: recommendation systems in algorithmic feeds and conversational language model systems (Section~\ref{sec:case-studies}). 

\section{Background \& Related Work}
\subsection{Theory of Affordance}
\label{subsec:theory-affordance}
In the study and practice of user experience design, the theory of affordance, first articulated in the human-centered design space by Norman in 1988, has become a central analytical tool  \cite{norman1988design, kremer2023design}. An affordance is defined as the properties of a system communicating to users the possible actions to complete with or upon the system. Affordances are distinct from \textit{features} and \textit{outcomes}, as affordances “mediate between the properties of an artifact (features) and what subjects do with the properties of an artifact (outcomes)” \cite[p.~2]{davis2016theorizing}. For example, a button on a website is a feature; its affordance is the suggestion to the user that it can be clicked, where the outcome may be submitting a form or closing a window.

More recently, Davis’ Mechanisms and Conditions framework of affordances (M\&C framework hereinafter) has shifted the central question in affordance theory from \textit{what} technologies afford to \textit{how} technologies afford, for what subjects, and under what circumstances \cite{davis2020artifacts}. The \textit{mechanisms} in the M\&C framework specify a continuum of how a technology \textit{requests}, \textit{demands}, \textit{encourages}, \textit{discourages}, \textit{refuses}, or \textit{allows} user action. The \textit{conditions} specify how different users (e.g., more technically literate vs less technically literate) or circumstances (e.g., high pressure vs low pressure environments) may lead to different experiences of affordances. This framework addresses two critiques of affordance theory: the binary mechanism of afford or not afford, and the universal subject and experience of affordances \cite{davis2023affordances, osiurak2017affordance}. 

The M\&C framework is well suited to operationalize affordances within today's sociotechnical landscape, notably when applying it to AI systems \cite{davis2023affordances, scarlett2019rethinking}. In our work, we use it specifically to focus on the design features and corresponding affordances of (1) AI-generated content (i.e., displays of model output), (2) transparency (i.e., displays of model understanding like explanations or performance metrics), and (3) interaction (i.e., elements of user engagement and feedback) \cite{liao2022designing, lai2021towards}.

\subsection{From Neutral Designs to Harmful Ones?}
Affordances aid in understanding the potential for designs to lead to harmful impacts. Design features are tools for designers to convey a system's affordances to users \cite{norman1988design}. In turn, as theorized by Mathur et al., design affordances shape the \textit{choice architecture}, or the way different action options are arranged, presented, and framed for users, thereby highlighting some of these options over others \cite{lockton2009choice, mathur2021makes}. Essentially, as design affordances influence user-system interactions, choice architectures strategically organize these possibilities to subtly guide user behavior. A harmful design feature is one that facilitates actions detrimental to users in the short or long term. For example, a design pattern can lead to deceptive affordances, compromising and harming user autonomy by tricking users into performing certain actions \cite{brenncke2023theory, mathur2021makes}.

In our work, we follow Chordia et al., Di Geronimo et al., and related studies that do not consider designer intentions in identifying a design feature as harmful \cite{di2020ui, chordia2023deceptive, gray2018dark}. We take this position as our focus is on the outcomes of interactions, irrespective of intentions. In other words, if a design feature causes negative impact then it should be considered a harmful design. We also take this position since, as Di Geronimo et al. note, understanding intentions of designers is subjective and difficult to discern \cite{di2020ui}.
 
\subsection{Conceptualizing Risks \& Harms in Sociotechnical Systems}
Research on the negative impact of AI systems has expanded in response to their growing deployment. This research broadly conceptualizes harm as the negative outcomes of ``entangled dynamics between design decisions, norms, and power''.~\cite[p.~2]{shelby2023sociotechnical} Researchers have focused on both identifying and categorizing existing harms as well as anticipatory \textit{risks} of various harms, including representational, economic, and social harms \cite{shelby2023sociotechnical,weidinger2022taxonomy}. Researchers have also developed measurement tools such as impact assessments, algorithmic audits, and various model evaluation approaches \cite{raji2020closing, metcalf2021algorithmic, solaiman2023evaluating, buccinca2023aha}. In this research, the assessment of AI systems' negative impact involves creating indicators (such as error rates or bias measurements) that represent potential harms, and that can be used can be used to affect certain regulatory or technical decisions \cite{metcalf2021algorithmic}.

Research on AI impact and harm increasingly draws on HCI methods to assess how human behaviors and interactions with AI systems complicate safe and responsible AI practices \cite{liao2022designing,araujo2020ai}. In HCI research, a significant area of study is the study of `dark patterns,' a term originally coined by Brignull, which directly examines the harms inflicted by digital interface designs \cite{brignull2015dark}. In this work, harm is understood as user-centric, interaction-focused, and conceptualized more narrowly as direct impact on individuals. Early dark patterns studies focused on privacy and financial harms caused by features such as intrusive pop-ups in cookie banners \cite{soe2020circumvention, gray2021dark}, deceptive pricing layouts in online shopping websites \cite{mathur2019dark, voigt2021dark}, and compulsive reward systems in video games \cite{zagal2013dark, goodstein2021cat}. More recent studies have broadened to other domains and different types of harm; some researchers investigate ``attention-capture" harms, showing that dark patterns that lead to addictive consumption are prevalent on social media platforms \cite{monge2023defining,mildner2023engaging}. Others examine dark patterns across internet-of-things devices, highlighting many instances of privacy harms \cite{kowalczyk2023understanding}.In our work, we extend these studies to analyze interface designs in AI applications with a focus on the harms they pose to individual welfare and the feedback loops of human-AI behavior that could make them distinct. 

\section{Scoping Review of Design Patterns, Affordances, and Harms in AI Interfaces}
\label{sec:review}
To better understand the characteristics, mechanisms, and impact of harmful designs in AI interfaces, we first conduct a scoping review of the relevant literature. A scoping review is a methodical survey of existing literature to map key trends, concepts, and gaps in a specific subject; such reviews are deemed particularly effective in clarifying the current state of knowledge on a topic and surfacing new priorities for future research \cite{shelby2023sociotechnical, arksey2005scoping}. 

\subsection{Overview of Methodology}
\subsubsection{Identification and selection of relevant articles}
We first identified relevant articles by searching the electronic database of the ACM Guide to the Computing Literature, which combines the traditional ACM Digital Library with conference proceedings, books, journals, and abstracts of popular publishers such as IEEE and Springer. \footnote{ACM Guide to the Computing Literature, \url{https://libraries.acm.org/digital-library/acm-guide-to-computing-literature}} Our initial search included entries of any content type published from January 1, 2000 to February 29, 2024.

Table \ref{tab:freq} contains our search terms, which we defined along two axes: (1) AI applications and (2) potentially harmful designs, following the terminology in the relevant literature. For AI applications, we included the general terms `AI' and `algorithm.' However, given the lack of definitional consensus around these terms, we also focused our search on three currently widespread AI systems: predictive, recommendation, and conversational systems. For the terms related to harmful designs, we adopted a broad umbrella of terms used by Roffarello et al. and Mathur et al. in their studies on harmful designs \cite{monge2023defining, mathur2021makes}. 

\begin{table}[h]
  \caption{Search terms used in the scoping review}
  \label{tab:freq}
  \begin{tabular}{cp{0.7\linewidth}} 
    \toprule
    Category & Search Terms \\
    \midrule
    Technology & \texttt{"AI" OR "algorithm*" OR "chatbot*" OR "conversational agent*" OR "recommend* system*" OR "predictive system*" OR "large language model*" OR "generative *" AND} \\
    Design & \texttt{"dark pattern*" OR "unethical design" OR "evil design" OR "manipulative design" OR "persuasive design" OR "deceptive design" OR "unethical interface" OR "evil interface" OR "manipulative interface" OR "persuasive interface" OR "deceptive interface"} \\
  \bottomrule
\end{tabular}
\end{table}

Articles were eligible for inclusion if they met all of the following criteria: 
\begin{itemize}
    \item discuss, directly or indirectly, harmful designs (i.e., interfaces which lead users to perform actions against their preferences, intentions, and best interest),
    \item discuss AI systems, algorithmic systems, or intelligent systems.

\end{itemize}

Articles were excluded if they met one or more of these criteria: 
\begin{itemize}
    \item discuss AI-driven deception and manipulation unrelated to interface designs,
    \item discuss using AI as an intervention (e.g., methods for automated dark patterns detection).
\end{itemize}

Overall, the initial search identified a total of 295 records. We first removed duplicates and analyzed whether the remaining articles met the inclusion and exclusion criteria through a full-text read-through leaving us with a final set of 17 records. Due to the interdisciplinary nature of the targeted literature, we also conducted a supplementary search leveraging existing knowledge and additional sources to add 10 additional articles. This resulted in a final set of 27 articles.   

\subsubsection{Data analysis and coding process}
We first recorded article metadata such as authors, title, abstract, publication type and year, and publication venue. Then, we followed Roffarello et al. in extracting information from each article along three main lines: (1) \textit{characteristics}, (2) \textit{mechanisms}, and (3) \textit{impact} of the discussed designs \cite{monge2023defining}. We define \textit{characteristics} as the form a design feature takes in a user interface; \textit{mechanisms} as the design affordances or how a design feature \textit{requests}, \textit{demands}, \textit{encourages}, \textit{discourages}, \textit{refuses}, or \textit{allows} different user actions; and \textit{impact} as the effect of these actions on various user behaviors as well as individual and societal outcomes (if discussed). To investigate the nature of adaptability in AI interfaces, we also recorded if studies address feedback, long-term effects, or changes in the impact of design affordances over time. We use this information to conduct a thematic analysis, which is a method to identify, analyze, and report patterns within data to interpret various aspects of a research topic \cite{clarke2017thematic}.

\subsection{Results}
Articles in our corpus discussed the following AI systems: voice assistants (2), decision support systems (2), conversational systems (e.g., with language models) (7), home robots (1), recommendation systems in health, social media, and streaming platforms (12), and other AI systems such as in safety technologies (3). We organized the findings into four main themes, each representing distinct types of harmful design patterns in AI interfaces: ``traditional" dark patterns, focusing on deceptive and manipulative interface elements; anthropomorphism, examining human-like elements and how they can mislead or misrepresent; explainability and transparency, addressing elements that communicate clarity and understanding; and, seamless design and lack of friction, highlighting elements that facilitate effortless yet impulsive user interaction. Under each theme, we analyze and discuss the characteristics, mechanisms, and impact of the relevant design features. We acknowledge that some of these design patterns also have positive aspects and serve critical functional purposes. However, in our review and discussion, we focus primarily on potential harms and do not discuss benefits as these are much more widely understood and discussed.

\subsubsection{“Traditional” dark patterns}
Multiple reviewed studies discuss design patterns in AI interfaces that have been previously categorized as dark patterns. These studies explicitly identify design features as ``dark patterns” or use the names of recognized dark pattern strategies, as articulated by Gray et al., such as \textit{obstruction} or \textit{sneaking} \cite{gray2018dark}. For example, Zhang et al. explore user privacy concerns with ChatGPT finding that \textit{obstruction} is employed to dissuade users from altering their data usage settings \cite{zhang2023s}; they show that to prevent chat data from being used in training future models, users must either lose access to the chat history feature or navigate a hidden request within an FAQ page to retain it---a task none of their study participants could accomplish. Alberts and Van Kleek examine phone notifications from social agents, showing that \textit{nagging}, a dark pattern involving persistent and intrusive prompts, is used to force users into ``social situations” with these agents, contributing to anthropomorphic perceptions \cite{alberts2023computers}. And, in interfaces with recommendation systems, several dark patterns were described, including those concerning time unawareness, such as hiding time on the screen, and extreme countdowns, such as autoplay timers \cite{chaudhary2022you}. Many of these traditional dark patterns in human-AI interfaces seem to serve data extraction purposes, either through increasing engagement or obfuscating privacy controls \cite{lacey2019cuteness,zhang2023s,leschanowsky2023privacy}.

\subsubsection{Anthropomorphism}
In human-AI interaction, anthropomorphism can be defined as attributing human traits, emotions, or intentions to AI systems \cite{salles2020anthropomorphism}. The anthropomorphism of AI systems by users is generally described as the outcome of anthropomorphic cues interacting with existing user mental models of these systems \cite{papagni2021pragmatic, hwang2022ai}. Anthropomorphic cues manifest in various forms. First, cues manifest in the appearance or visualization of systems. For example, humanoid representations may lead to attributions of sentience and an overassignment of agency \cite{hwang2022ai, papagni2021pragmatic}. In contrast, Lacey and Caudwell identify “cuteness as a dark pattern” and how it can result in attributing powerlessness to systems \cite{lacey2019cuteness}. Second, cues may be embedded in interfaces in the form of “priming statements” that humanize AI agents. Pataranutaporn et al. find that priming statements portraying an AI system's motive as good, evil, or neutral, shape user mental models and interactions, even when the AI model itself remains unchanged \cite{pataranutaporn2023influencing}. The effect they observe is even more pronounced with generative models. Finally, cues can be present in the natural language output of AI systems, like conversational or language models, where they may create ambiguity between social and transactional interactions. This may occur by incorporating language that references social relations, feelings, and emotions, effectively blurring the distinction between a user interacting with an AI system and with a human \cite{zhan2023there,clark2019makes, abercrombie2023mirages}. 

Anthropomorphism directly affects how users perceive AI systems, leading them to ascribe specific intentions and motives to these systems. The impact of these perceptions on user actions is well-established in cognitive science and was also present in several articles \cite{nanay2013between, bannon1995human, dahlstedt2018action, wen2022sense}. For example, anthropomorphic cues in AI systems can foster a sense of trust among users. This aspect is explored by Zhang et al. and Lacey and Caudwell, who illustrate how such trust, partly derived from anthropomorphic features, can be manipulated to serve third-party interests or conceal data collection, leading to unintended sensitive disclosures and privacy harms \cite{lacey2019cuteness,zhang2023s}. Additionally, some articles explain how anthropomorphism can evoke emotional responses in users, where using characteristics resembling those of real individuals, such as their voice, can establish even stronger emotional connections \cite{lacey2019cuteness,chan2021kinvoices}. On the other hand, anthropomorphic cues may lead to an “uncanny valley” effect, resulting in algorithmic anxiety or fear \cite{zhan2023there,papagni2021pragmatic}. Thus, design choices that lead to various anthropomorphic perceptions shape the extent of users' trust and attachment in interactions with AI systems.

\subsubsection{Explainability and transparency}
The presence or absence of an explanation is itself a consequential design decision. In examining social media platforms, Brady et al. discuss how not labeling content as algorithmically recommended may influence user engagement \cite{brady2023algorithm,nelimarkka2019re}. On the other hand, research on trust and overreliance shows that including AI explanations is positively associated with the capability of an AI system, regardless of the content of the explanation \cite{liao2022designing}. The presence or absence of explanations can drive unearned trust and flawed mental models of systems’ capabilities, capacity to err, and potential biases \cite{chromik2019dark,mucha2021interfaces}. Finally, in the context of transparency in AI disclosures, research suggests that when AI systems are used but their usage is not disclosed, anthropomorphic perceptions of those systems increase \cite{papagni2021pragmatic,hwang2022ai}. 

However, research indicates that even when explanations are provided, there are two potential ways in which they can be deceptive: (1) the way explanations are phrased (e.g., with a level of confidence not matching the actual level of certainty associated with the model output or the explanation), or (2) the way explanations are presented (e.g., visualized) in the interface. On the latter, Mucha et al. discuss how ``selective highlighting," or using visual cues to emphasize certain parts of an explanation while neglecting others \cite{chromik2019dark}, could conceal explanations that point to unethical AI predictions. Chromik et al. argue that another issue is burying explanations deep within multiple layers of settings menus, which they also describe as a dark pattern \cite{mucha2021interfaces}. In that way, design decisions around AI explanations influence how useful, deceptive, and impactful they are. 

\subsubsection{Seamless design and lack of friction}
Typically, user interfaces exhibit frictionlessness through seamless designs. For example, in online platforms, such designs include features like autoplay, infinite scrolling, and the absence of context around presented content \cite{masrani2023slowing, chaudhary2022you, brady2023algorithm, kender2022shape, albuquerque2023social}. While seamless designs can streamline user experience and positively enhance it, they may also inadvertently obscure critical reflection and informed decision-making. In the context of AI systems, the literature on recommendation systems in online platforms frequently cites the lack of friction caused by seamless designs on these platforms as a source of mismatch between user preferences and user actions. The concept of frictionlessness is often discussed in relation to `system 1’ and `system 2’ behaviors, terms initially coined by Kahneman \cite{kahneman2011thinking}. Frictionlessness tends to encourage system 1 behaviors, which research in psychology and cognition describes as automatic and reactive (i.e.,\textit{ thinking fast}), as opposed to system 2 behaviors, which are more intentional and reflective (i.e., \textit{thinking slow}), aligning more closely with people’s true preferences \cite{kahneman2011thinking,masrani2023slowing,sharevski2021voxpop}. 

For example, Agan et al.'s audit of Facebook's recommendation algorithm shows that automatic behavior can lead to out-group human, and consequently, algorithmic discrimination beyond explicit preferences \cite{agan2023automating}. Further, research discusses how frictionlessnes might curtail mental deliberation in online media consumption, amplifying human biases towards moral-emotional content and potentially contributing to societal phenomena like echo chambers and misinformation spread \cite{chen2023spread, sharevski2021voxpop}. Frictionlessness was also associated with attentional harms and regrets due to its facilitation of mindless and compulsive consumption \cite{chaudhary2022you,schaffner2023don,masrani2023slowing}. Importantly, while there may be user controls that increase friction (e.g., disabling autoplay), studies found that they are limited and, when present, buried in setting menus, making them inaccessible to the majority of users \cite{schaffner2023don,chaudhary2022you}.

\subsubsection{Temporal effects}
About 25\% of the articles in our corpus address the long-term effects of the phenomena they examine on user behavior. These brief references primarily focus on the role of reinforcement learning in shaping human behavior \cite{brady2023algorithm,zhan2023there}, the formation of habits \cite{chaudhary2022you}, and the evolution of user mental models of AI systems over time \cite{papagni2021pragmatic}, sometimes identifying feedback loops \cite{pataranutaporn2023influencing}. Thus, while alluding to their importance, the articles do not thoroughly interrogate temporal effects neither theoretically nor empirically. Consequently, their assessment of impact was limited; while they could describe the immediate impact of affordances on individual behavior, they could only speculate about the long-term or societal effects. 

\section{DECAI: Design-Enhanced Control of AI Systems}
\label{sec:decai}
Our scoping review identified four salient categories of harmful designs in AI interfaces, and suggested that there is limited research addressing long-term effects caused by feedback loops of human-AI behavior. In this section, we build upon these findings to support the evaluation of potential negative impact from AI interface design features, such as those discussed in our scoping review. We introduce a model, which we call DECAI, that borrows from principles of control systems theory to represent repeated human-AI interactions.

\subsection{DECAI's Contributions} 
\subsubsection{Generating hypotheses for iterative and empirical assessments} 
Current impact assessment frameworks prioritize the identification, evaluation, and mitigation of harms associated with the use of AI systems \cite{metcalf2021algorithmic,buccinca2023aha}. DECAI aims to facilitate the early stages of these assessments, specifically targeting interface designs. It can assist practitioners in systematically identifying particular design features for detailed examination and provides a structured approach for formulating focused, testable hypotheses suitable for empirical study. DECAI also implicitly considers the diverse conditions of users, such as their technical proficiency and emotional states, in that process ensuring that hypothesis generation takes various user experiences into account. 

\subsubsection{Motivating the utility of control systems theory}
We draw on control systems theory due to both its pragmatic and conceptual relevance. First, this field of research emphasizes the dynamic study and control of multi-component systems that are inherently variable, mirroring the sociotechnicality and adaptability of human-AI systems \cite{anand2013introduction}. Second, it highlights control and moderation within systems---two themes conceptually connected to autonomy-undermining designs and research on AI harm and deceptive designs more broadly~\cite{lyshevski2001control, boyd2022designing, birhane2022values, gray2021end}. Increasingly, HCI research has grappled with the challenges of designing user experiences for AI systems particularly due to (1) the uncertainty of AI systems' capabilities and (2) the complexity of AI systems' outputs, which may range from simple to adaptively complex \cite{yang2020re, dove2017ux, yang2016planning}. This context of variability and complexity underscores the relevance of DECAI’s approach, which integrates control systems theory to respond to the dynamic nature of sociotechnical systems and their often overlooked feedback loops.

DECAI models a core property of modern AI systems: their \textit{adaptability}. Although other properties (e.g., \textit{stochasticity}, \textit{agenticness}) may also be significant to designing interfaces (see Section ~\ref{sec:discussion}), we concentrate on adaptability as it is both a widespread property and one that directly contributes to the evolution of the human-AI system through feedback mechanisms \cite{reader2022models, matias2023humans}. In the following sections, we introduce the system components of DECAI (\Cref{subsec:system-components}), detailing its control objectives (\Cref{subsec:control-objective}) and the nature of its inputs and outputs (\Cref{subsec:inputs-outputs}). \Cref{subsec:cycle-interaction,subsec:temporal-evolution} propose five distinct stages to evaluate the impact of a design feature on user behavior and welfare during cycles of human-AI interaction.

\subsection{System Components}
\label{subsec:system-components}
In a typical control system, the \textit{controller} is the central decision-making component of the system. The \textit{process} is the entity being regulated by the controller. The \textit{actuator} is the component that implements the controller output, and the \textit{sensor} is the component that monitors and relays new input on the process state back to the controller. In response, the controller adjusts its control strategy and consequently its future output \cite{nise2010control}. For example, in a home heating system, the thermostat acts as the \textit{controller}, regulating the room's temperature, which is the \textit{process}. The heating or cooling unit, the \textit{actuator}, adjusts the temperature based on the thermostat's settings. The thermostat's built-in \textit{sensor} monitors the room temperature and feeds this information back to the controller, enabling the thermostat to continually fine-tune its temperature settings over time \cite{haines2006control, nise2010control}.

\subsubsection{DECAI components}
\begin{figure*}
\centering
\includegraphics[width=0.9\textwidth]{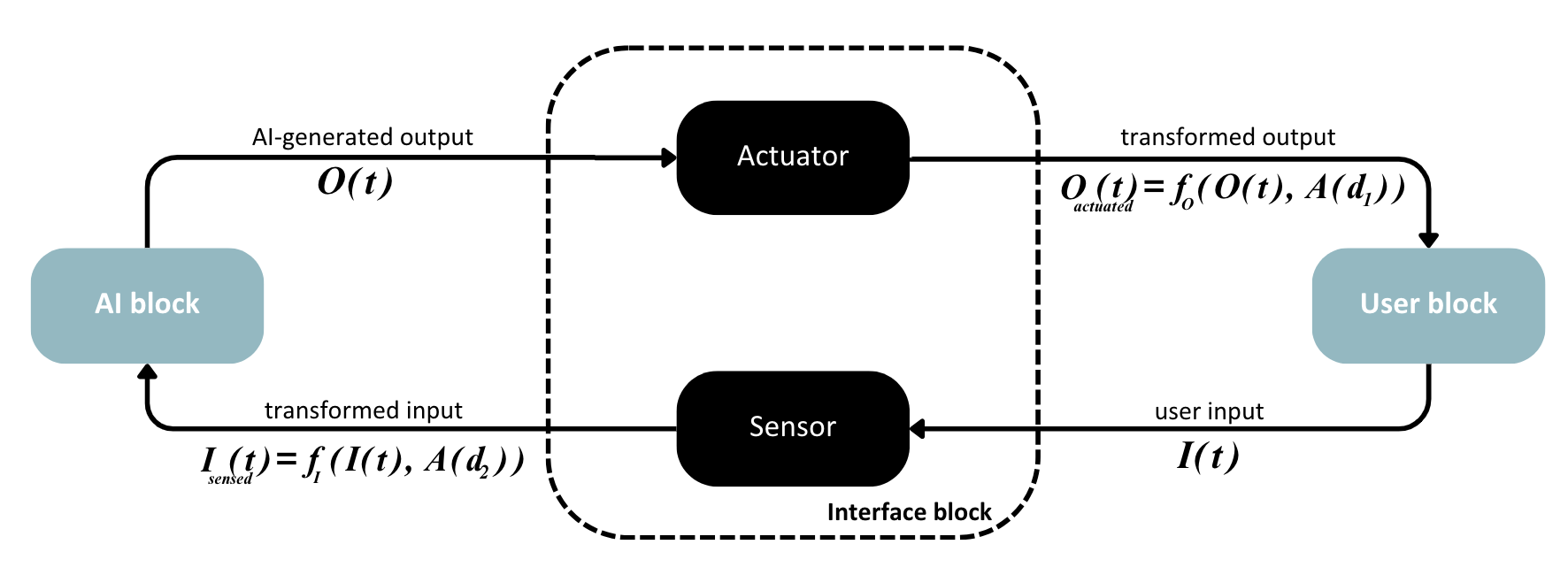}
\caption{DECAI model consisting of an AI block, user block, and interface block.}
\label{fig:figure-block}
\end{figure*}

Figure ~\ref{fig:figure-block} presents a block diagram of our proposed closed-loop system model. We model the \textbf{AI block} as the controller and the \textbf{user block} as the process. We model the \textbf{interface block} as including both the actuator and the sensor. The actuator transforms AI-generated output into a format that can be presented to the user in the interface. The sensor gathers new input data from the user's interactions with the system's output, and relays this data back to the controller, creating a feedback mechanism \cite{franklin2002feedback}. While our DECAI is informed by control systems theory, it diverges from a traditional directional control model: instead of the AI block exerting control over the user block, our approach emphasizes the collaborative interaction of these two components toward achieving the control objective \cite{lewis2013cooperative}.

\subsubsection{Initializing DECAI} At time $t=0$, we identify the user input available to the AI block as limited to one or a combination of three options: (1) \textbf{manual} setting of user preferences, such as directly selecting interests when onboarded to a new social media application; (2) \textbf{third-party} information, such as cookies and location data, as a proxy for user preferences; or (3) no user-specific input, in which case the system is set to a predefined \textbf{default} setting. At $t > 0$, the interface changes to reflect any AI-generated output based on this initial set of user preference data (manual, third-party, or default), and the sensor begins to collect new user input in response to these outputs. For example, on a streaming platform, the recommendation algorithm (AI block) generates content recommendations for the user (user block) based on their initially set interests. 

\subsection{Control Objective}
\label{subsec:control-objective}
In complex and real-world systems, optimizing for and balancing multiple control objectives is often necessary \cite{dujardin2015multi, brandao2019multi}. For instance, in a home heating system, along with maintaining the room temperature, the system may need to optimize for energy efficiency and minimize costs. To simplify DECAI and its objectives, we limit it to human-AI systems consisting of one individual and one AI system. As our goal is to assist those investigating the negative impact of interface designs, we suggest the primary control objective as the minimization of user harm. This objective aligns with widely recognized goals among technology providers, policymakers, and researchers to minimize harm in interactions with advanced AI systems \cite{EuropeanCommission2019TrustworthyAI,IEEE2019EthicallyAlignedDesign, green2021contestation}. 

\subsection{Inputs and Outputs}
\label{subsec:inputs-outputs}
Table ~\ref{tab:decai} displays a full list of DECAI's variables, functions, and their definitions. The interface block in our model clarifies the dual functionality embedded in human-AI interfaces, which both present AI-generated output and collect user input. We define the system's user input as $I$ and AI-generated output as $O$. The constraints, defined as the limitations that restrict the possible values, on $I$ and $O$ arise from two sources. First, there are \textbf{\textit{hard} constraints}, determined by an AI system’s inherent technical limitations. For example, a conversational AI system may be limited to textual data processing, and thus users may only enter text input and receive text output. Second, there are \textbf{\textit{soft} constraints}, shaped by a human-AI system's design features. These features influence the presentation of AI-generated output and the nature of user input in the system. Our model is focused on studying how the affordances of design features constitute soft constraints.

To model the impact of affordances, we utilize the action intensities defined in the M\&C framework (see \Cref{subsec:theory-affordance}). These action intensities range from \textit{demand}, representing the strongest push for a certain user action, to \textit{refuse}, representing the strongest opposition to a certain user action. In between, there exist intermediate intensities such as \textit{encourage} for moderate promotion of an action, \textit{request} for a mild suggestion, \textit{allow} for neutral permission, and \textit{discourage} for moderate dissuasion. A user action can be defined as an action taken by the user either on the system (e.g., in the form of user input) or outside of a system (e.g., in the form of an external decision). An affordance of a design feature $d$ consists of both an action intensity and a user action, and can be represented as $A(d)$.

\begin{table}[h]
\centering
  \caption{Taxonomy of DECAI variables with an example on recommendation systems}
  \label{tab:decai}
    \begin{tabular}{p{0.1\columnwidth}p{0.3\columnwidth}p{0.5\columnwidth}}
    \toprule
    Variable & Definition & Example of a video recommendation feed \\
    \midrule
    $t$ & Time & Daily updates\\
    $I(t)$ & User input at time $t$ &  Engagement signal (user likes a video)\\
    $O(t)$ & AI-generated output at time $t$ & Generated video recommendations \\
    $f_{\text{AI}}$ & Maps user input to AI output & Updates video recommendations based on engagement signal (user likes a video)\\
    $C(user)$ & User condition & Low or high technical proficiency \\
    $d$ & Design feature & Hidden or visible `dislike' button\\
    $A(d,t)$ & Affordance of a design feature $d$ at time $t$ & The button encourages or discourages signalling disinterest \\
    $f_{\text{O}}$ & Presents output to user & Presents video recommendations with appropriate metadata in user feed \\
    $f_{\text{I}}$ & Collects input from user & Collects engagement signal (user likes a video) \\
  \bottomrule
\end{tabular}
\end{table}

\subsection{The Cycle of Human-AI Interaction}
\label{subsec:cycle-interaction}
As the human-AI interaction cycle commences at $t=0$ with the AI system generating its first output, \(O\), we propose the following three stages for investigating the impact of an interface design feature in a single cycle:

\paragraph{Stage 1---What are the conditions of the receiving user?} A user processes AI-generated output and provides input according to their individual conditions, $C(user)$. Drawing on both the conditions from the M\&C framework and the empirical results of our scoping review, we propose considering five condition axes: (1) cognitive ability, (2) technical proficiency, (3) domain expertise, (4) context of use (e.g., physical and environmental constraints), and (5) psychological or emotional state of the user \cite{davis2016theorizing,liao2022designing}.

\paragraph{\textit{Stage 2---What are the relevant interface design features and their affordances?}} The design features of both the AI-generated output and the user input should be identified: 
\begin{enumerate}
    \item \textit{How is the AI-generated output presented in the interface?} As seen in Figure ~\ref{fig:figure-block}, the user does not receive the raw AI-generated output, \(O\), but rather a processed version from the actuator, \(O_{\text{actuated}}\), which is designed to be comprehensible and useful to the user. This is shaped by both the content of the output and one (or several) interface design features, denoted as \(d_1\). $d_1$ may have one or a set of affordances, $A(d_1)$, that may influence how a user receives this transformed output. This results in  $O_{\text{actuated}} = f_O(O, A(d_1))$. 
    \item \textit{How do different user preferences translate to user action on the interface?} Then, the user may respond to this output by providing new input, \( I \), to the interface. However, the input action is influenced by the design affordances, \( A(d_2) \), of one (or several) design features, denoted as $d_2$, in the interface at the point of data collection by the sensor. This results in \( I_{\text{sensed}} \), where $I_{\text{sensed}} = f_I(I, A(d_2)) $

\end{enumerate}

\paragraph{\textit{Stage 3---What is the impact of these affordances on the user state?}}
Considering the user's conditions, the intensity and associated user action of the affordances should be mapped to their potential impact on user behavior and welfare.

This single cycle ends with the AI block receiving the new user input, \( I_{\text{sensed}} \).

\subsection{Evolution Over Time: Repeated Cycles \& Feedback Loops}
\label{subsec:temporal-evolution}
The cycle described above repeats numerous times and the interaction dynamics between the user and AI system evolve via continuous feedback loops. As the AI block processes the new user input, \( I_{\text{sensed}}(t) \), from the sensor, which may reflect updated user preferences or needs, it produces an updated output, \( O(t+1) \). This results in a feedback loop where $ O(t+1) = f_{\text{AI}}(O(t), \alpha \cdot I_{\text{sensed}}(t))$, and where $\alpha$ represents a scaling coefficient that determines how much the user's input affects subsequent AI-generated output and $O(t)$ represents the previous latest AI-generated output. 

We propose examining two types of feedback: (1) \textit{reinforcing feedback} or positive feedback, which amplifies behaviors or patterns in the system, and (2) \textit{optimizing feedback} or negative feedback, which adjusts the system to more closely align with the initial control objective \cite{matias2022impact, aastrom2021feedback}. 

\paragraph{\textit{Stage 4---How does the impact of these design affordances evolve over time?}}
We suggest that the impact of interface designs on human-AI interactions is specifically influenced by \textit{reinforcing feedback}, as it shapes the long-term evolution of user behaviors in response to the AI system. To model this, we introduce a time-dependent affordance function, $A(d,t)$. This function captures how the intensity and associated user action of a specific design affordance evolves, either increasing or decreasing over time. Consider a case where there is no external intervention or shift in the user condition, $C(user)$, within the control system. In this case, the user's mental models and the AI system’s behavior reinforce each other over time, increasing the action intensity of an affordance in its original direction. For instance, a design $d$ that initially \textit{encourages} a user action might gradually shift towards \textit{demanding} that action as $t$ increases, at least for a specified duration. 

\paragraph{\textit{Stage 5---What is the frequency of these updates?}}
Parties with access to more information about the AI system, like system developers, may also analyze the rhythm of this interaction cycle. The frequencies of output presentation by the actuator, data collection by the sensor, and data processing by the AI system, can vary. This variability means the actuator and sensor may operate on different cycles, affecting the nature and timing of the feedback loop \cite{aastrom2021feedback}. 

\section{Case Studies}
\label{sec:case-studies}
Here, we show how to use DECAI to interrogate design choices in AI interfaces. We illustrate its usage with two case studies drawn from the results of our scoping review. The case studies offer testable hypotheses and ground our model in a range of user conditions, input and output affordances, and categories of harm.

\subsection{Recommendation Systems in Algorithmic Feeds}
Research on recommendation systems in online streaming platforms has shown that these systems could lead users to polarizing and conspiratorial content \cite{ribeiro2020auditing,hosseinmardi2021examining}. More recently, the rise of primarily algorithmically-curated feeds (or ‘For You’ feeds) has particularly intensified interest in the impact of these feeds on online information consumption. Much of the resulting journalism and research has focused on how these systems lead young users down ‘rabbit holes’ of harmful mental health content \cite{Amnesty_International_2023,harwood2021tiktok,bahnwegeffects}. Here, we show how DECAI could be used to study the design features of these feeds and their consequences on young users.

\paragraph{Stage 1---What are the conditions of the receiving user?} With a focus on young users, we identify two main user conditions, $C_1(user)$ and $C_2(user)$, and hypothesize about how they influence the reception of recommended content. $C_1(user)$ is the \textbf{psychological and emotional state} of young users; the majority of young users are more emotionally impressionable and sensitive to social validation, and thus may gravitate towards content that resonates with their emotional state and peer group norms \cite{gwon2018concept, o2018social}. $C_2(user)$ is the \textbf{domain expertise} of young users; the majority of young users are limited in their domain expertise due to their age and exposure, and thus may rely more on surface-level features, such as source popularity, when evaluating content.

\paragraph{\textit{Stage 2---What are the relevant interface design features and their affordances?}} We identify the system output as the recommended content, and the system input as user feedback and engagement signals (e.g., likes, reposts, and watch time). 

Drawing on our scoping review (Section ~\ref{sec:review}), we first identify important design features and their affordances in presenting recommended content in vertical video feeds. The first feature, $d_1$, is the \textbf{low-friction autoplay and infinite scroll} feature \cite{chaudhary2022you, kender2022shape}. It has three affordances: $A_1(d_1)$ \textit{discourages} intentional consumption, $A_2(d_1)$ \textit{encourages} users to spend more time on the application, and $A_3(d_1)$ \textit{allows} for passive engagement through views and watch time, rather than explicit signals of interest \cite{schaffner2023don}. The second feature, $d_2$, is the \textbf{lack of \textit{in-situ} explanation of recommendation type}, such as minimal or non-existent ‘sponsored’ labels \cite{kender2022shape}. It has one affordance: $A(d_2)$ \textit{discourages} users from assessing the incentives of the content creator.

Then, we identify important design features and their affordances in collecting user input to be relayed back to the recommendation system. The first feature we identify, $d_3$, is the \textbf{limited accessibility of the ‘dislike’ or ‘not interested’ button}; for example, on TikTok, this button is not readily visible alongside other vertical engagement buttons but requires a user to either click ‘share’ or long press on the screen to access it. This design has one affordance: $A(d_3)$ \textit{discourages} users from providing an explicit signal of disinterest. Another feature, $d_4$, is the \textbf{suggestion of search terms} related to the topics of the recommended content, whose affordance, $A(d_4)$, \textit{encourages} users to further explore the topics in the already recommended content using these suggested search terms.

From these features, we choose to assess the impact of the limited accessibility of the `dislike' or `not interested' button and its affordances.

\paragraph{\textit{Stage 3---What is the impact of these affordances on the user state? }} Here, we hypothesize about the impact of the button design affordances on young users. First, addressing its impact on user behavior, we hypothesize that most users are unable to locate the dislike button and thus do not use it. Then, we address potential harms from these user behaviors; we speculate that when users are exposed to and view harmful content, even if they recognize it as harmful, they are unable to express their active disinterest in the topic. Their passive viewing of this content is interpreted as interest, as most popular recommendation systems factor in watch time as an engagement signal, leading to similar future recommendations \cite{yi2014beyond}.

\paragraph{\textit{Stage 4--- How does the impact of these design affordances evolve over time?}} Finally, we could examine how these design affordances evolve in their impact on user engagement with recommended content over time. This would clarify how designs contribute to the development of rabbit holes, as that phenomenon occurs gradually, rather than immediately. We can examine the potential reduction in the action intensity of some affordances. For example, we can ask: do users reach a point in the interaction loop where the action intensity of an affordance decreases, for instance, from \textit{discourage} to \textit{allow}? In this case, under what conditions and at what point do users who find themselves in a rabbit hole decide to actively search for and use the `not interested' button? Or, we can address potential increases in action intensity and ask: at what point does infinite scrolling lead to addictive consumption, \textit{demanding} rather than \textit{encouraging} viewing more content \cite{chaudhary2022you}?

\subsection{Conversational Language Model Systems}
Large language models (LLMs) are being increasingly integrated into conversational user interfaces. As models continue to generate plausible but inaccurate information, concerns of potential overreliance on LLM-generated output have grown \cite{leiser2023chatgpt, maynez2020faithfulness}. Here, we analyze how design choices in `typical’ conversational LLM interfaces (i.e., interfaces with text-input fields and linear conversation flow displays) contribute to flawed mental models of LLMs, especially when users are seeking high-stakes specialized advice, such as medical advice.  

\paragraph{\textit{Stage 1---What are the conditions of the receiving user?}} We identify two user conditions, $C_1(user)$ and $C_2(user)$, and hypothesize about how they affect users’ mental models of LLMs in conversational interfaces. $C_1(user)$ is users' \textbf{technology proficiency}; the majority of lay users will have limited technical knowledge of how LLMs work, and thus may overestimate LLMs’ capabilities and underestimate their limitations \cite{zou2023exploring}. $C_2(user)$ is users' \textbf{domain expertise}; the majority of lay users seeking medical advice from an LLM will have limited domain knowledge, and thus may be less effective at discerning the quality of LLM responses.

\paragraph{\textit{Stage 2---What are the relevant interface design features and their affordances?}} We identify the system output as the LLM-generated text presented to the user, and the system input as user queries entered in the input text box. 

We first identify some design features and their affordances in presenting the LLM-generated output. 
The first set of features we identify are \textbf{anthropomorphic cues}, $d_1$, such as humanoid profile pictures, typing indicators, and natural language cues. Such cues have several affordances: $A_1(d_1)$ \textit{allows} for social and emotional engagement, $A_2(d_1)$ \textit{allows} for developing human-like trust, and $A_3(d_1)$ \textit{encourages} sensitive disclosures \cite{hwang2022ai,zhang2023s}. The second feature, $d_2$, is the \textbf{lack of or insufficient (e.g., hidden in an about page) disclosure} of LLM use. The following are affordances of this lack of transparency: $A_1(d_2)$ \textit{encourages} sensitive disclosures and $A_2(d_2)$ \textit{discourages} fact-checking \cite{papagni2021pragmatic}.

Then, we examine important design features and their affordances in collecting queries from users. The main feature we identify is the \textbf{inability to edit input} after submission, $d_3$. This feature has two affordances: $A_1(d_3)$ \textit{refuses} revision and $A_2(d_3)$ \textit{discourages} sensitive disclosures. 

From these features, we choose to assess the impact of anthropomorphic cues and their affordances.

\paragraph{\textit{Stage 3---What is the impact of these affordances on the user state? }} Here, we hypothesize about the impact of these design affordances on users. We first address impact on user behavior, and hypothesize that as users inquire about medical advice, they reveal sensitive information, such as personal identification data and medical history. If they receive an inaccurate response from the system, they are less likely to critically evaluate this response and more likely to accept it, assigning it human-like trust \cite{pataranutaporn2023influencing}. Then, we address potential harms from these user behaviors: since some companies may engage in user data collection to improve future models or fine-tune existing ones, there are data leakage and profiling risks \cite{plant2022you}. Additionally, depending on the nature of the medical advice, overreliance on the advice presented may translate into real-world action or inaction that could harm the user.

\paragraph{\textit{Stage 4---How does the impact of these design affordances evolve over time?}} Finally, we could examine how user interactions with these cues evolve, focusing on the development and reinforcement of user mental models of LLMs. We can first examine the potential reduction in the action intensity of some affordances: when users are repeatedly exposed to anthropomorphic LLMs, does their growing familiarity lead them to perceive the LLMs as less social and more mechanistic \cite{papagni2021pragmatic}? Then, we can address potential increases in action intensity and ask: do users, over time, reinforce their anthropomorphization of LLMs as a result of natural language anthropomorphic cues in the LLM output \cite{zhang2023s}?

\section{Discussion}
\label{sec:discussion}
We view our work as an initial step towards rigorously examining the reality of AI interfaces as critical sites for harm propagation and reduction. Through a scoping review, we organize and present four main categories of harmful AI interfaces, namely focused on dark patterns, anthropomorphism, explainability and transparency, and seamless design and lack of friction. We then develop a conceptual model, DECAI, to structure and facilitate investigations into the impact of AI interface designs, such as those we categorize, or to discover new ones. Both our model and its case studies highlight the need to examine the cascading impact of human-AI feedback loops influenced by interface designs.

\textbf{Empirical and iterative testing with DECAI.} Through two case studies, we showed how DECAI can be used to develop hypotheses for evaluating design-mediated human-AI interactions. Real-world impact assessments are often more challenging, due to the need to carefully isolate effects, select features and affordances, and operationalize hypotheses, as well as due to the costs required to systematically examine long-term impact. Thus, we recommend using DECAI to support empirical, iterative evaluation of design patterns. For example, we recommend identifying design features and affordances through heuristic reviews or think-aloud user studies \cite{mildner2023engaging, alhadreti2018rethinking}. Then, we then recommend testing of the hypotheses generated through DECAI by conducting online experiments and longitudinal user engagement studies to empirically examine the impact of design features on user decision-making and well-being \cite{kjaerup2021longitudinal}.  

\textbf{Regulatory opportunities.} Audits of dark patterns have gained notable regulatory traction, being codified into both EU and US law~\cite{dsa2023, dma2023, dataactproposal2023}. We believe that our work can aid further research and evidence collection building on this momentum in the AI context \cite{king2021regulating}. Currently, AI-focused legislation such as the EU AI Act only indirectly addresses interface designs through regulations on transparency, human oversight, and AI-driven manipulation \cite{EUAIACT2023}. However, interface designs may allow for the circumvention of such regulation, for example, by allowing for the strategic placement of transparency disclosures to diminish their visibility. Our results also point to how harmful designs can be used to further data collection, a practice central to today’s AI industry \cite{Kak_West_2023}. In that way, interface designs can be pivotal in shaping both current and future regulations aimed at safe and responsible AI practices.

\textbf{Beyond adaptive AI systems.}
Our work focuses on the adaptability of AI systems and the feedback loops of human-AI behavior at play. However, design challenges associated with other properties of AI systems must also be examined in future research. For instance, the \textit{stochasticity} of LLMs has amplified concerns around predictability, explainability, and accuracy \cite{zhao2023explainability, yang2020re}. Similarly, the increasing \textit{agenticness} of AI systems has drawn attention to harms associated with systems capable of pursuing complex goals with limited human supervision \cite{chan2023harms}. For the latter, seamless interfaces may be especially damaging, as friction could facilitate interruptibility and prevent unwanted outcomes \cite{shavitpractices}. More research is needed to unravel the design complexity of interfaces attending to various such properties of AI systems.

\textbf{Beyond digital interfaces.} Additionally, while digital interfaces are the subject of the vast majority of the literature, the adoption of modern AI systems in interactive \textit{physical} interfaces is likely to increase \cite{Lambert_2023}. Interacting with AI systems in physical space brings about its own set of design challenges around accessibility, data collection, and deception \cite{owens2022exploring}. As physical interfaces may increase the possible design affordances and their risks, it is important to further investigate and expand impact assessment models like DECAI to accommodate these considerations \cite{hartson2003cognitive}.

\textbf{Considering disparate impact.} DECAI considers user conditions across several axes, including knowledge and well-being, in its impact assessment approach. We made this choice in response to research consistently demonstrating that computing harms do not affect all users equally \cite{shelby2023sociotechnical, aizenberg2020designing}. For instance, in the context of human-AI interactions, vulnerable and marginalized groups, such as children, the elderly, and those with limited technical literacy, may have more flawed mental models of AI systems, their capabilities, and their risks. We believe that DECAI is an initial step that researchers can build upon to better evaluate how interface designs may disproportionately impact different societal groups and harm \textit{collective} rather than \textit{individual} welfare. 
 
\begin{acks}
Lujain Ibrahim acknowledges funding from the Oxford Internet Institute - Dieter Schwarz Foundation Research Program and The Mozilla Foundation.
\end{acks}

\bibliographystyle{ACM-Reference-Format}
\bibliography{citations}
\appendix
\end{document}